\documentclass{emulateapj}
\newcommand     \etl    {et al.}
\shortauthors{Bai \etl}

\begin{document}
\title{Search for IR Emission from Intracluster Dust in A2029}
\author{Lei\,Bai\altaffilmark{1},
George\,H.~Rieke\altaffilmark{1},
Marcia\,J.~Rieke\altaffilmark{1}}
\email{bail@as.arizona.edu}
\altaffiltext{1}{Steward Observatory, University of Arizona, 933 N. Cherry Avenue, Tucson, AZ 85721}
\begin{abstract}
We have searched for IR emission from the intracluster dust (ICD) in the galaxy cluster A2029. 
Weak signals of enhanced extended emission in the cluster are detected at 24 and 70 \micron.
However, the signals are indistinguishable from the foreground fluctuations.
The 24 versus 70 \micron\ color map does not discriminate the dust emission in the cluster from the cirrus emission.
After excluding the contamination from the point sources, we obtain upper limits for the extended ICD emission in A2029, $5\times 10^3~ \rm{Jy~ sr^{-1}}$ at 24 \micron\ and $5 \times 10^4~ \rm{Jy~ sr^{-1}}$ at 70 \micron. 
The upper limits are generally consistent with the expectation from theoretical calculations and support a dust deficiency in the cluster compared to the ISM in our galaxy.
Our results suggest that even with the much improved sensitivity of current IR telescopes, a clear detection of the IR emission from ICD may be difficult due to cirrus noise.

\end{abstract}
\section{INTRODUCTION}
An important constituent of the X-ray-emitting gas in massive clusters is thought to be material lost from cluster galaxies. 
Such gas would be well processed and contain a significant component of heavy elements, indicated by the absorption lines in the X-ray spectra.
This galactic-originated gas would also introduce dust grains into the intracluster (IC) gas, which can survive sputtering by ambient gas for a typical time scale of $\sim 10^8$ yr.
However, in contrast to the heavy elements, solid evidence of the presence of dust grains in IC gas is still lacking.
The indirect measurement of ICD, suggested by enhanced visual extinction towards clusters deduced from high-redshift objects, is still very controversial \citep[e.g.,][]{Maoz95}.

A direct measurement of ICD can be provided by its IR emission when they are collisionally heated by hot X-ray emitting gas.
\citet{Dwek90} modeled this process and predict flux levels of about $0.2~ \rm{MJy~sr^{-1}}$ at 100 \micron\ in the Coma cluster. 
Their calculations, and the upper limit constraints provided by $IRAS$ suggest a deficiency of dust in the central region by a factor of 100 compared with the interstellar dust-to-gas ratio.  
Using ISO data, \citet{Stickel98} report the detection of a flux excess of $\sim0.1~\rm{MJy~sr^{-1}}$ at 120 \micron\ in the central region of the Coma cluster and attribute this excess to the ICD emission.
\citet{Stickel02} confirmed this excess in the $I_{120~\micron} / I_{180~\micron}$ color profile across the Coma core region, but they did not find the same signal in the other 5 clusters they studied.
\citet{Quillen99} estimated the flux in Coma that might arise from faint infrared-excess galaxies and concluded that it could be a significant fraction of the apparent extended component.
\citet{Bai06} have studied the Coma cluster and found the projected number density of infrared-emitting galaxies is the highest in the core region. 
Such an excess of IR galaxies could cause an excess of IR emission in the cluster core region, raising more questions about the origin of the excess emission reported by \citet{Stickel98, Stickel02}.

The observational situation is therefore muddled. 
The difficulties are mostly associated with the sensitivity and detector stability in previous missions, but also with their limited angular resolution which has made it difficult to distinguish the composite of the emission from individual cluster galaxies from that of ICD.
The low resolution also causes confusion limitations due to extragalactic sources.
But with the much-improved sensitivity and spatial resolution of current and future IR telescopes, e.g., Spitzer, AKARI and Herschel, we may marginally reach the detection thresholds predicted by theoretical calculations \citep{Yamada05}.

In this paper, we present an attempt to measure the extended IR emission from ICD in the galaxy cluster A2029 at $z=0.077$. 
It is one of the brightest X-ray clusters nearby, with a X-ray luminosity $L_{X}(2-10 {\rm keV})=1.1\times10^{45} h_{70}^{-2}~ {\rm ergs~ s^{-1}}$ \citep{David93} and an intracluster gas metallicity about a third of the solar metallicity \citep{White00}. 
The cluster is in a region of relatively low IR cirrus, which makes it a perfect target for searching for the ICD emission.
The size of its X-ray emitting gas ($\sim 2 \arcmin$), which limits the size of the ICD emission, is well suited to sensitive measurements with Spitzer. 
\citet{Hansen00} attempted to assess the IR emission from ICD in this cluster with ISO data, but with no success. 
Using MIPS observations, we can disentangle the IR emission of the galaxies from that of the ICD in A2029 and are able to set upper limits to the ICD emission at 24 and 70 \micron. 
Throughout the paper, we assume a $\Lambda$CDM cosmology with parameter set $(h,\Omega_{0},\Lambda_{0}) = (0.7,0.3,0.7)$.

\section{\sc {Observations and data reduction}}
The cluster was first observed in Feb. 2004 with MIPS in a medium rate scan mode.
The 24 \micron , 70 \micron\ and 160 \micron\ data were acquired simultaneously for the central $40\arcmin \times 60\arcmin$ region.
The integration time was about 80, 40, and 8 seconds at 24, 70 and 160 \micron.
To further confirm the weak extended emission detected in the cluster core region, a second observation was obtained in Jan, 2005 to tripled the observation time.

The data were processed with the MIPS Data Analysis Tool \citep[DAT version 2.9;][]{Gordon05}. 
At 24 \micron, extra steps were carried out to remove scattered light, which appears as a low-level background modulation synchronous with scan mirror position.
We generated average backgrounds for each scan mirror position and subtracted them from the images.
The final mosaic combining old and new observations has an exposure time of $\sim$ 340 s pixel$^{-1}$ on average.
For 70 \micron, we only use the data from the second observation because the first observation was taken before the bias was changed on the 70 \micron\ array and has quite noticeable stim latents.
The total exposure time for the 70 \micron\ image is $\sim$ 120 s pixel$^{-1}$ and is $\sim$ 30 s pixel$^{-1}$ for the 160 \micron\ image.
At the 70 and 160 \micron\ bands, we took extra processing steps to improve the images.
The pixel-dependent backgrounds were subtracted from the images, which were obtained by fitting the images outside of the central $6\arcmin$ region of the cluster.
This process tends to de-emphasize the extended emission outside the source region, but it keeps the signal we are interested in.
The mosaics have point-spread functions (PSF) with FWHMs of $\sim6\arcsec,~18\arcsec, ~40\arcsec$ at 24, 70 and 160 \micron.
 
To avoid contamination from individual galaxies, we extract all the IR sources from the images using Sextractor.
For the 24 \micron\ image, we exclude all the pixels contributing to IR sources down to the $1~\sigma$ detection limit ($\sim45~\mu Jy$). 
After the source extraction, 74\% of the pixels remain in the 24 \micron\ image.
For the 70 and 160 \micron\ data, due to the lower uniformity of the mosaic image, we only exclude the IR sources down to the $2~\sigma$ ($\sim 7$ mJy and $\sim 50$ mJy, respectively) sensitivity level, after which, 76\% and 85\% of the pixels remain at each band. 
If the IR source density in the cluster core region were greatly enhanced by the cluster galaxies, this source extraction process might cause a bias when we try to measure the extended emission from the remaining pixels.
However, this is not the case.
The fractions of the remaining pixels in the central $r<2\arcmin$ region of the cluster are 70\%, 80\% and 90\% at 24, 70 and 160 \micron, respectively, which do not differ significantly from the average values for the whole images.

\section{RESULTS}
After excluding the IR sources, we smooth the images using a median filter with a box size of $\sim 2\farcm 5$. 
The filter matches roughly the size of the X-ray emitting region \citep[to $\sim$ 10\% of the central gas density,][]{Lewis03}.
All IR images show significant brightness fluctuations across the map, and the patterns of the fluctuations are roughly consistent at the three wavelengths.
The fluctuations are probably due to the foreground cirrus emission, but they could also be caused by the fluctuation of the number density of unresolved faint sources, especially at 70 \micron, where the observations have not reached the confusion limit due to extragalactic sources \citep{Dole04} and the cirrus emission is still relatively weak.

\begin{figure}
\centering
\includegraphics[totalheight=2in]{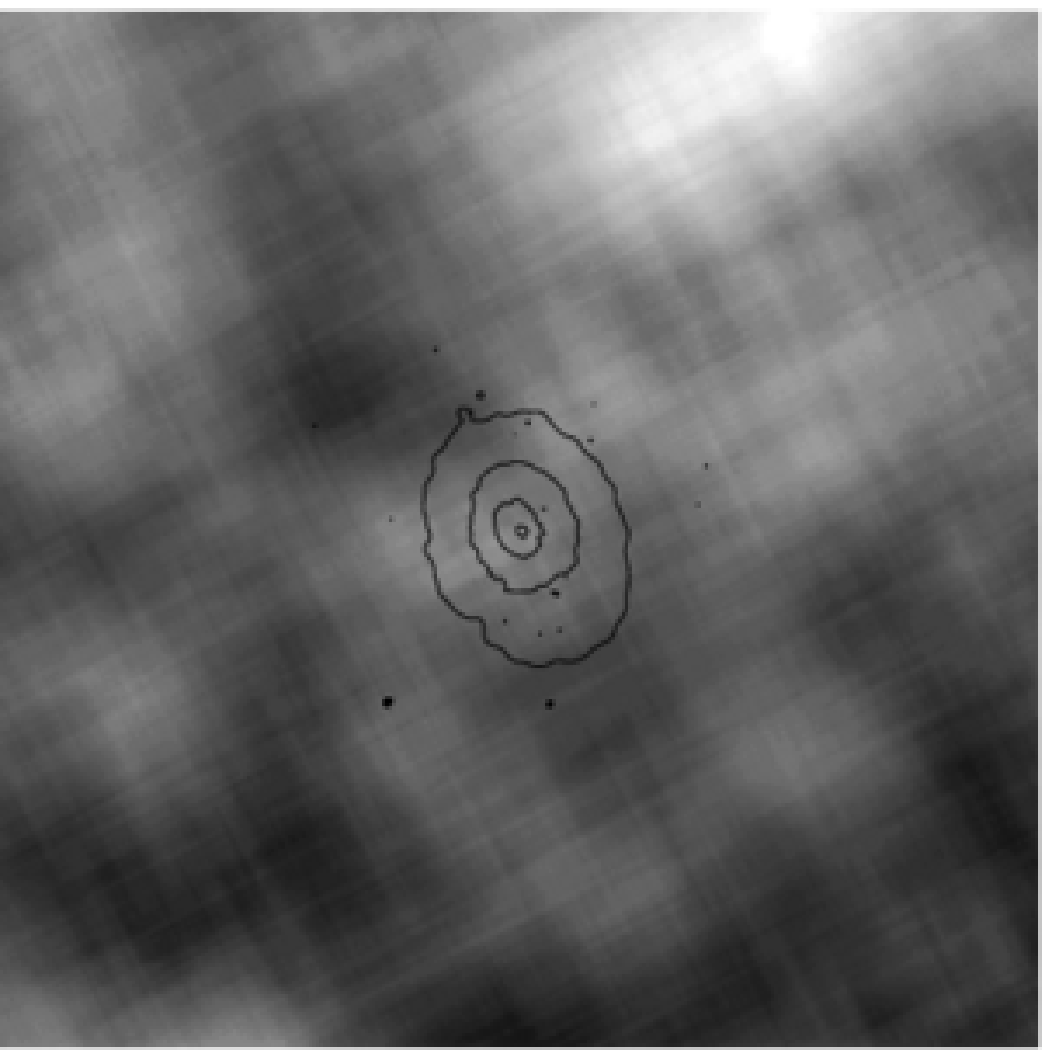}
\includegraphics[totalheight=2in]{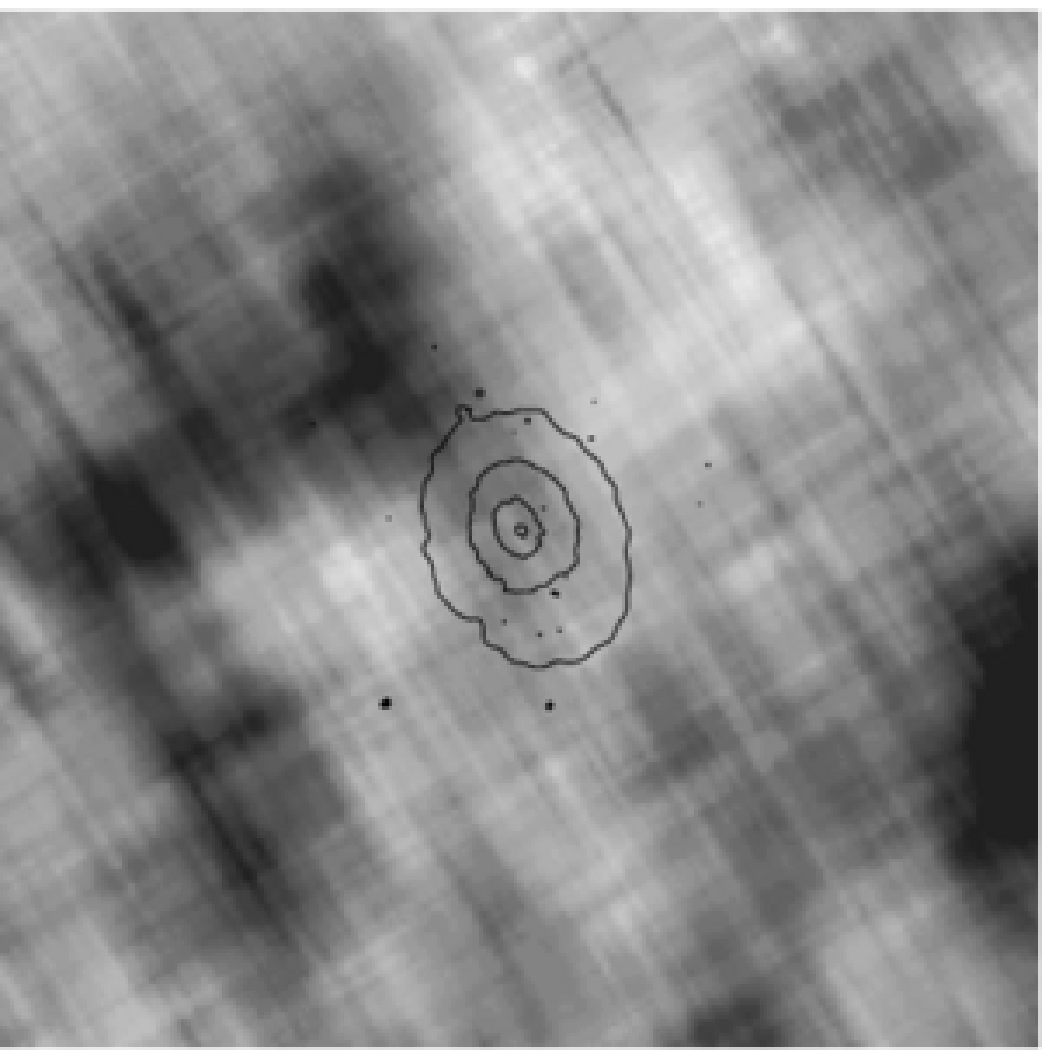}
\includegraphics[totalheight=2in]{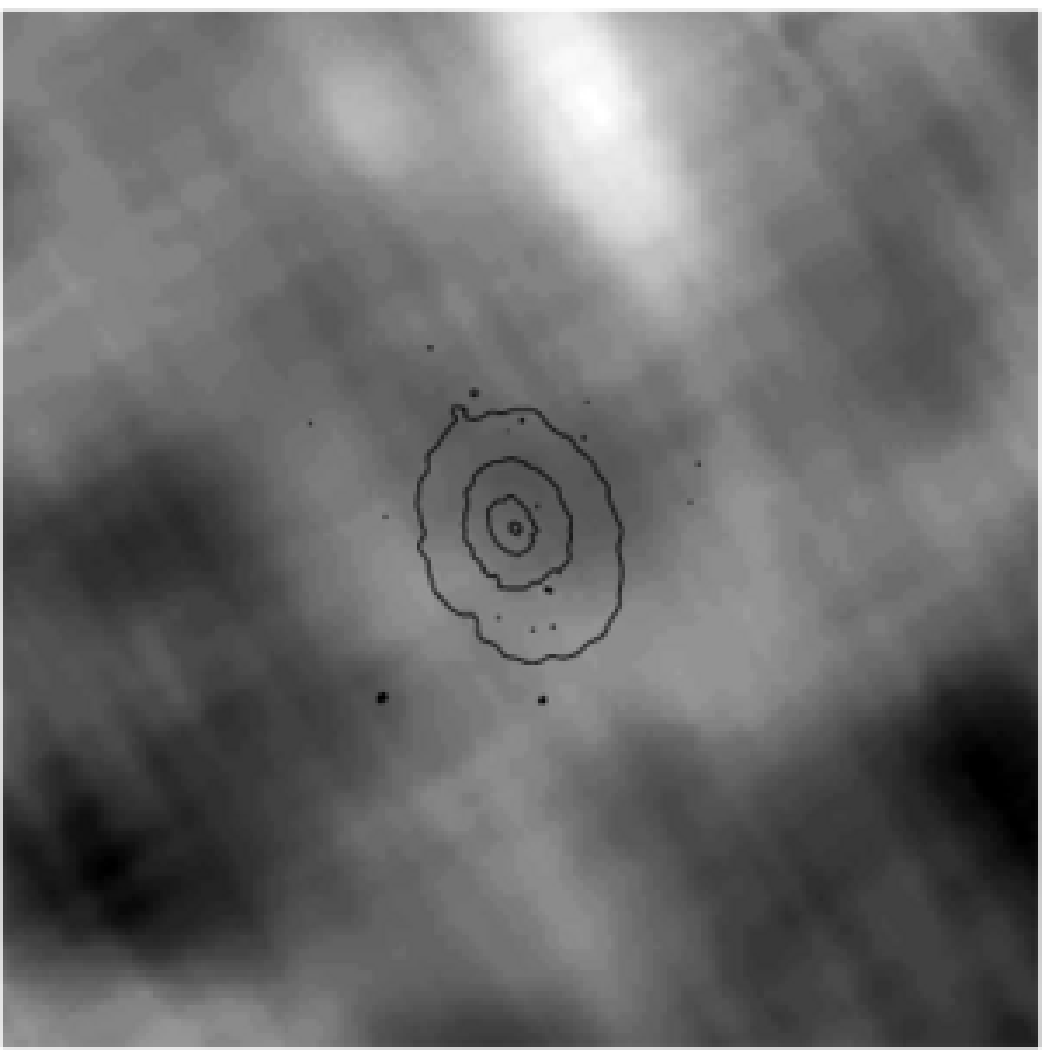}
\caption{24 \micron, 70 \micron\ and 160 \micron\ images smoothed with a median filter of size $\sim 2\farcm 5$.
The overlaid X-ray contours are logarithmic, and spaced by 0.48.
The image sizes are $20\arcmin \times 20\arcmin$.  East is to the left and north is at the top.
The low-level grid patterns are artifacts of the boxcar smoothing.}
\label{f_img}
\end{figure}

In Fig.~\ref{f_img}, we show the smoothed images with the X-ray contours overlaid.
The X-ray data has an exposure time of 79 ks and were retrieved from the Chandra Data Archive (ObsID 4977). 
The 24 \micron\ smoothed image shows a weak enhancement corresponding to the X-ray emitting region, but it is indistinguishable from the foreground cirrus structure and its morphology has no apparent similarity with the X-ray image.
At 70 \micron, the enhancement is even harder to recognize due to a large cloud of faint emission crossing the cluster.
The core region is also slightly fainter than the outer region of the X-ray emitting region.
The enhancement, if any, is likely just to be the extended emission from the Galactic cirrus.
At 160 \micron, the X-ray emitting region is totally indistinguishable from the background.

\begin{figure}
\centering
\includegraphics[totalheight=2in]{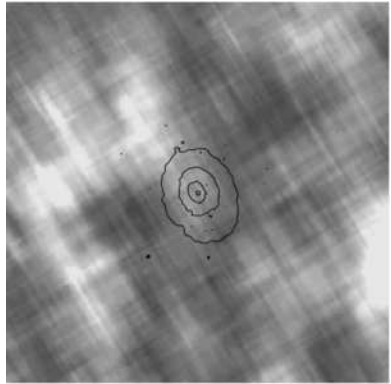}
\caption{The surface brightness map of $I_{24~\micron}/I_{70~\micron}$, with X-ray contours overlaid.  The size and direction of the image, as well as the scale of the contours, are the same as in Fig.~\ref{f_img_color}.}
\label{f_img_color}
\end{figure}

Using ISO observations, \cite{Stickel98,Stickel02} found a peak in the $I_{120~\micron}/I_{180~\micron}$ surface brightness ratio towards the Coma cluster center and they use this color difference to discriminate the ICD emission from the foreground cirrus.
To see if a similar enhancement of the brightness ratio occurs in A2029, we derived a color map by dividing the 24 \micron\ image by the 70 \micron\ image.
Since the backgrounds have already been subtracted from both images and we are mostly interested in comparing the color of the cluster emission with that of the interstellar cirrus, we add the average brightness of the interstellar cirrus estimated by SPOT \footnote{SPOT, the Spitzer tool for planning observations and submitting proposals, http://ssc.spitzer.caltech.edu/propkit/spot/} before we calculate the brightness ratio.
In Fig.~\ref{f_img_color}, the X-ray emission region shows a warmer color than the faint cloud shown in the 70 \micron\ map.
However, again, this enhancement is not significant compared to the fluctuations in the color map, and there are some other patches in the image showing similar color to the X-ray emission region. 
We do not show the $I_{70~\micron}/I_{160~\micron}$ color map here because the color is dominated by the fluctuations at 70 \micron\ due to the much larger cirrus emission at 160 \micron\ than at 70 \micron.  

\begin{figure}
\plottwo{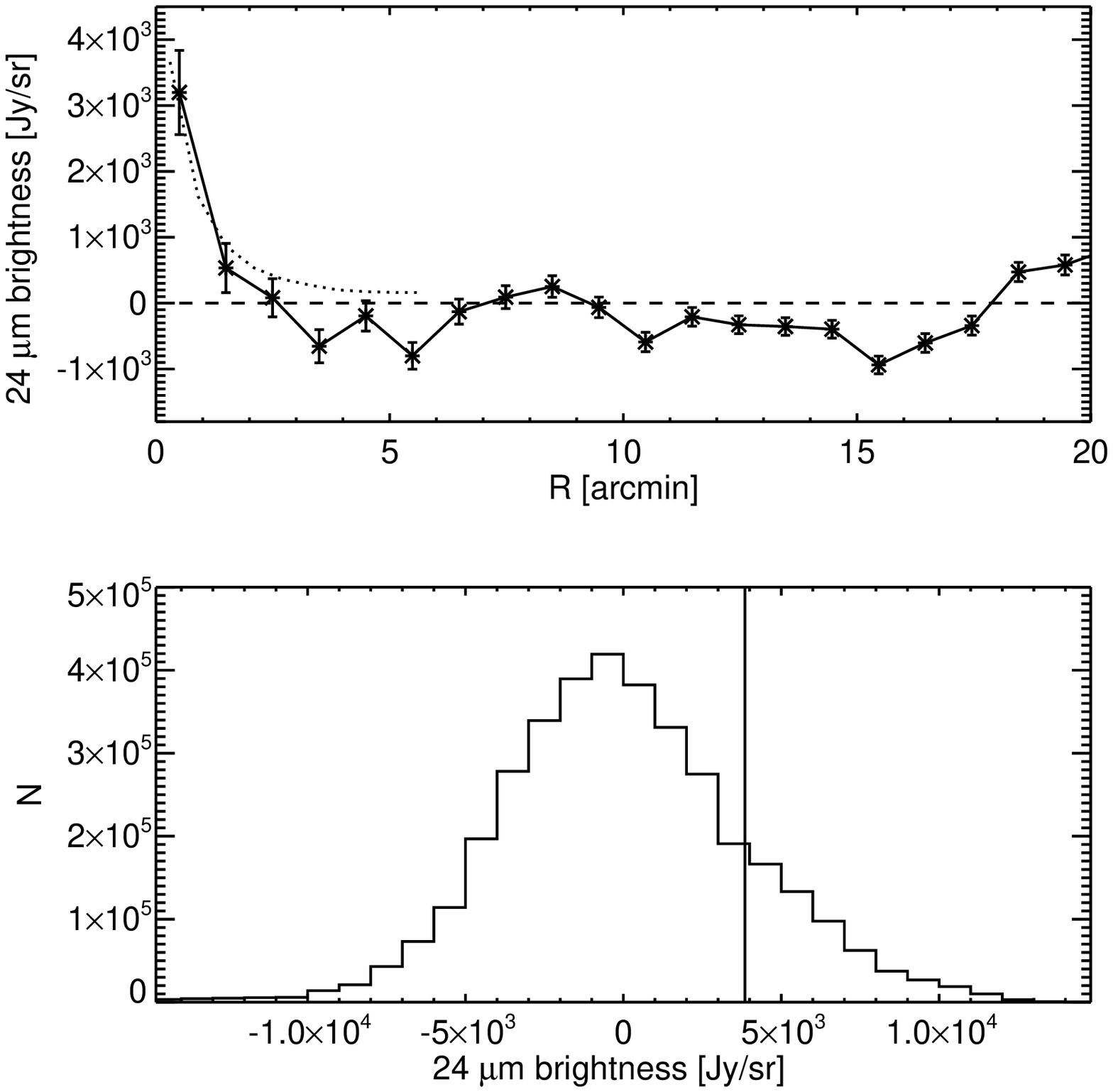}{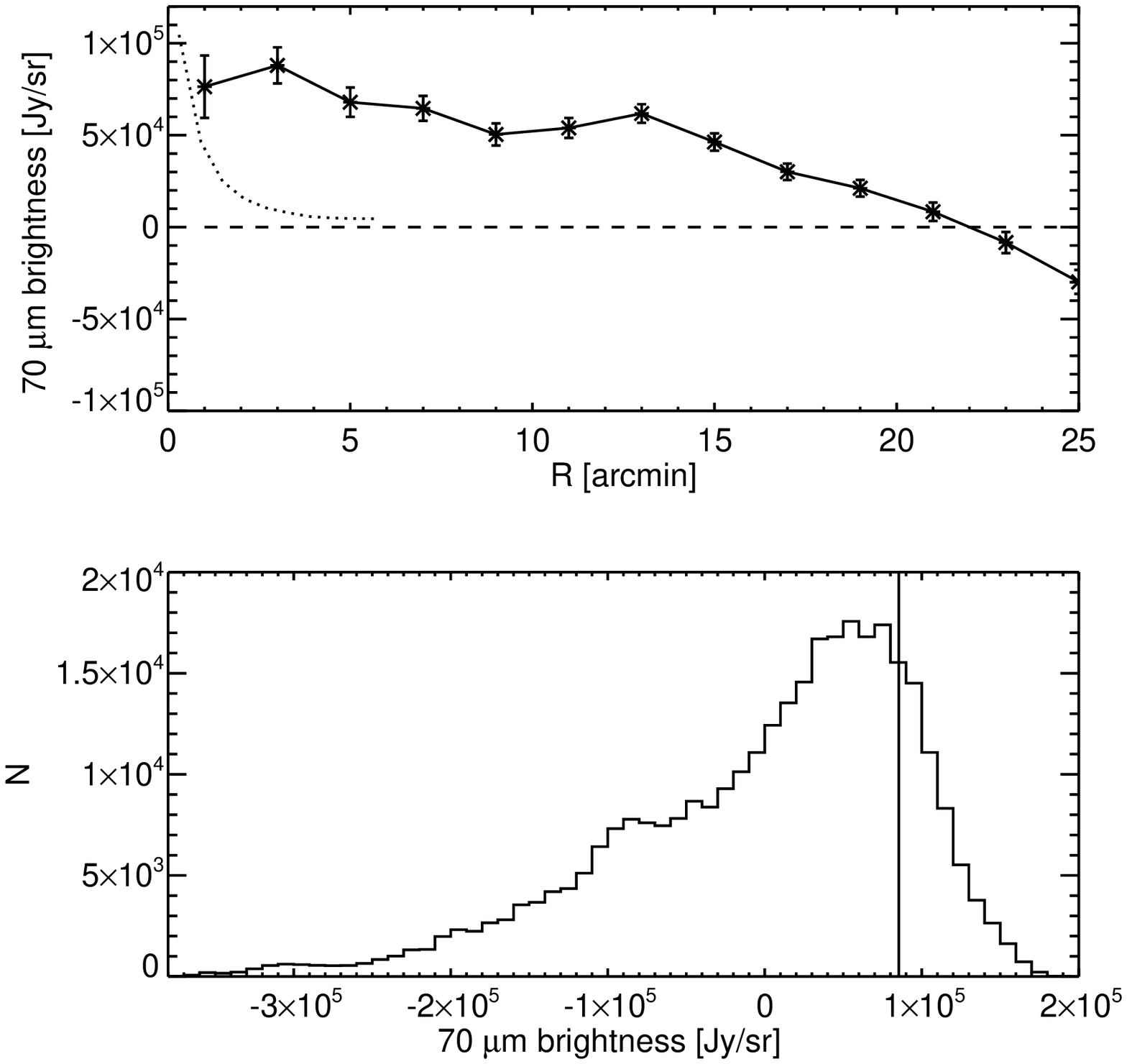}
\caption{Top: the radial profiles of brightness at 24 and 70 \micron\ as a function of distance from the center of the cluster.  The error bars are the errors of the mean in each region.
The dashed lines are the average values for each image.
The X-ray brightness profiles are plotted as the dotted lines.
Bottom: the histograms of the smoothed images. 
The 24 \micron\ image is smoothed with a median box of size $80 \arcsec$ and the 70 \micron\ image is smoothed with a median box of size $160 \arcsec$.
The vertical lines are the values at the center of the cluster.
}
\label{f_radprof}
\end{figure}

Although the extended IR emission of the cluster is indistinguishable from the foreground cirrus in the images, we can obtain an upper limit to the ICD emission.
In Fig.~\ref{f_radprof}, we plot the average surface brightness as a function of distance from the cluster center, derived from the point sources-removed, non-smoothed image.
The error bars are the standard error of the mean for each data point.
The 24 \micron\ radial profile shows a clear peak at radius $<1\arcmin$, coincident with the peak of the X-ray brightness profile.
The enhancement within $r<1\arcmin$ relative to the average of the whole image sets a $2\sigma$ upper limit of $5\times 10^3~ \rm{Jy~ sr^{-1}}$ for the extended emission from the ICD.
However, the apparent significance of the peak in the radial profile is mostly due to the fact that at large radii, the surface brightness is the average of pixels in a larger area, in which the small scale flux density fluctuations have already been canceled out.
To show the fluctuation of the brightness at a smaller scale, we plot the histogram of the pixels in the IR sources masked-out, smoothed 24 \micron\ image. 
The smoothing uses a median box of size $80\arcsec$; it retains the fluctuations at spatial scales larger than the box size. 
The surface brightness at the center of the cluster, although also showing an enhancement ($\sim4\times 10^3~ \rm{Jy~ sr^{-1}}$), is only $1\sigma$ higher than the average of the smoothed image.

At 70 \micron, the enhancement in the cluster is less obvious due to the foreground cloud it lies in.
The peak of the emission is not at the center of the cluster, but about $2\arcmin$ away. 
Therefore we estimate the enhancement in the central $r<2\arcmin$ region, and obtain a $2\sigma$ upper limit of $1\times 10^5~ \rm{Jy~ sr^{-1}}$.
However, if we exclude the emission in the cluster coming from this foreground cloud and compare the enhancement of the cluster relative to the local average brightness in the $r<15\arcmin$ region (approximately the scale of the cloud), we have a smaller upper limit, $5 \times 10^4~ \rm{Jy~ sr^{-1}}$.
The cloud also causes the histogram of the smoothed image (with a median box of size $160\arcsec$) to peak slightly off the mean.   
Although the surface brightness at the cluster center shows an enhancement at about $0.9\sigma$ relative to the average, the enhancement is less significant compared to the peak of the distribution.
We can also compare our results with the average brightness fluctuation of cirrus emission given by \citet{Miville07}, which is about $2 \times 10^4~ \rm{Jy~ sr^{-1}}$ at this scale for a low brightness region.
The upper limit we measured is only a few times larger than the average cirrus fluctuation, which again proves the insignificance of the enhancement compared to cirrus noise.
At 160 \micron, the cluster region shows a lower than average surface density compared to the whole image.
The $2\sigma$ error of this non-detection is $6\times 10^4~ \rm{Jy~ sr^{-1}}$.

\section{DISCUSSION AND CONCLUSIONS}
\citet{Dwek90} calculated the infrared emission of the ICD collisionally heated by X-ray-emitting hot gas in the Coma cluster, which has a similar temperature and metallicity, but a higher density than the hot gas in A2029 \citep{White00}.
Therefore the expected dust emission in the Coma cluster should be larger than that in A2029 and can provide an upper limit to the dust emission in A2029.
In Coma, at a distance of about 100 kpc from the cluster center, the expected dust emission is about $5\times10^2\rm{~Jy~sr^{-1}}$ at 25 \micron\ and $3\times10^4\rm{~Jy~sr^{-1}}$ at 60 \micron.
The upper limit we obtained at 24 \micron, $5 \times 10^3~ \rm{Jy~ sr^{-1}}$ within $\sim$90 kpc of A2029, is 10 times larger than the expected intensity.
At 70 \micron,  the upper limit excluding emission from the foreground cloud, $5\times10^4~\rm{Jy~sr^{-1}}$, which does not change much from $r< 90$ kpc to $r<180$ kpc, is very close to the expected intensity.
According to \citet{Dwek90}, their calculation suggests the dust in the central region of a cluster is deficient by a factor of 100 relative to the standard value for the ISM. 
The agreement between our upper limit measurements and their calculation confirms that the dust, if any, should be at least deficient by that amount.

\begin{figure}
\epsscale{0.5}
\plotone{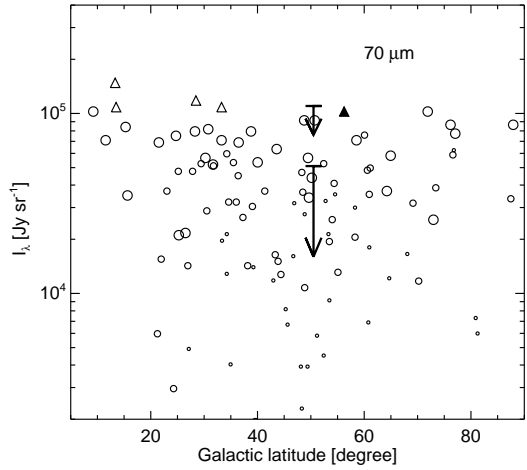}
\caption{Expected 70 \micron\ intensities of ICD in the central 175 kpc of a sample of clusters from \citet{Yamada05} compared with the upper limits measured in A2029 in the same region. 
The triangles are for the five clusters with highest 70 \micron\ intensity.
The filled triangle is A3112, which we expect to have a similar level of emission as A2029.
The large circles are clusters at redshift $0.01<z<0.1$, medium circles are at $0.1<z<0.3$ and small circles are at $0.3<z<0.8$.  
The top downward arrow is the 2$\sigma$ upper limit for A2029 measured relative to the whole 70 \micron\ image; the lower downward arrow is the upper limit obtained by subtracting the local foreground emission ($r<15 \arcmin$). 
}
\label{f_yamada}
\end{figure}

\citet{Yamada05} performed a comprehensive study of the emission of ICD in a sample of clusters with $z\sim 0.01-0.8$.
They present the expected 70 \micron\ intensities at the projected distance of 20 kpc from the center of the clusters (Fig.~7 of their paper).
In order to compare their results with the upper limits we obtained for A2029, we scale the intensities at 20 kpc with a factor of 0.28 to deduce the average intensities in the central 175 kpc region, the same region in which we measured the upper limits for A2029.  
The factor is obtained by assuming the clusters have a similar intensity radial profile as illustrated by the Perseus Cluster in their paper.
The results are shown in Fig.~\ref{f_yamada}.
The two upper limits we obtained are plotted as the downward arrows. 
They are very close to the predicted intensities of the clusters at similar redshift (the large circles and triangles).
In particular, there is one cluster, A3112 (shown as the filled triangle), with very similar properties to A2029.
They both are regular, X-ray bright clusters with redshift $z\sim0.07$ and at similar Galactic latitudes.
According to the X-ray analysis by \citet{White00}, on which \citet{Yamada05} base their calculation, A2029 has very similar metallicity, gas temperature and density compared to A3112, which would indicate an IR ICD emission at the same level as that in A3112.
The expected intensity in A3112 is very close to the conservative upper limit we measured but is higher than the upper limit that excludes the emission from the foreground cirrus.
The discrepancy suggests that the theoretical calculation by \citet{Yamada05} may overestimate the observed ICD emission. 

Another way to detect the weak IR emission from ICD is by stacking many galaxy clusters.
\citet{Montier05} stacked a large sample (11,507) of $IRAS$ cluster maps together and found a statistical detection of $2.1\pm0.7 \times 10^3~\rm{Jy~sr^{-1}}$ at 25 \micron\ and $1.6\pm0.7 \times 10^4~\rm{Jy~sr^{-1}}$ at 60 \micron, both several times smaller than the upper limits we obtained in A2029.
This is consistent with the fact that A2029 is one of the brightest X-ray clusters and we would expect the ICD emission in this cluster to be higher than the average value of a large sample of clusters.

The faint average value from stacking plus our failure to detect significant signals in a cluster expected to be relatively bright together indicate that the IR emission of ICD gas must be generally very faint. 
Even with the much improved sensitivities of the current generation of IR telescopes, this faint ICD emission will be very hard to be distinguished from cirrus fluctuations.

\acknowledgments
This work was supported by funding for \textit{Spitzer} GTO programs by NASA, through the Jet Propulsion Laboratory subcontracts \#1255094 and \#1256318.
We thank K. Gordan, M. Blaylock, M. Block, and J. L. Hinz for help in data reduction and the referee for comments.

\bibliographystyle{apj}

\begin{thebibliography}{}
\bibitem[Bai et al.(2006)]{Bai06} Bai, L., Rieke, G.~H., Rieke, M.~J., Hinz, J.~L., Kelly, D.~M., \& Blaylock, M.\ 2006, \apj, 639, 827
\bibitem[David et al.(1993)]{David93} David, L.~P., Slyz, A., Jones, C., Forman, W., Vrtilek, S.~D., \& Arnaud, K.~A.\ 1993, \apj, 412, 479
\bibitem[Dole et al.(2004)]{Dole04} Dole, H., et al.\ 2004, \apjs, 154, 93
\bibitem[Dwek et al.(1990)]{Dwek90} Dwek, E., Rephaeli, Y., \& Mather, J.~C.\ 1990, \apj, 350, 104 
\bibitem[Gordon et al.(2005)]{Gordon05} Gordon, K.~D., et al.\ 2005, \pasp, 117, 503
\bibitem[Hansen et al.(2000)]{Hansen00} Hansen, L., J{\o}rgensen, H.~E., N{\o}rgaard-Nielsen, H.~U., Pedersen, K., Goudfrooij, P., \& Linden-V{\o}rnle, M.~J.~D.\ 2000, \aap, 362, 133 
\bibitem[Lewis et al.(2003)]{Lewis03} Lewis, A.~D., Buote, D.~A., \& Stocke, J.~T.\ 2003, \apj, 586, 135
\bibitem[Maoz(1995)]{Maoz95} Maoz, D.\ 1995, \apjl, 455, L115
\bibitem[Miville-Desch{\^e}nes et al.(2007)]{Miville07} Miville-Desch{\^e}nes, M.-A., Lagache, G., Boulanger, F., \& Puget, J.-L.\ 2007, \aap, 469, 595 
\bibitem[Montier \& Giard(2005)]{Montier05} Montier, L.~A., \& Giard, M.\ 2005, \aap, 439, 35
\bibitem[Quillen et al.(1999)]{Quillen99} Quillen, A.~C., Rieke, G.~H., Rieke, M.~J., Caldwell, N., \& Engelbracht, C.~W.\ 1999, \apj, 518, 632 
\bibitem[Stickel et al.(2002)]{Stickel02} Stickel, M., Klaas, U., Lemke, D., \& Mattila, K.\ 2002, \aap, 383, 367
\bibitem[Stickel et al.(1998)]{Stickel98} Stickel, M., Lemke, D., Mattila, K., Haikala, L.~K., \& Haas, M.\ 1998, \aap, 329, 55 
\bibitem[White(2000)]{White00} White, D.~A.\ 2000, \mnras, 312, 663
\bibitem[Yamada \& Kitayama(2005)]{Yamada05} Yamada, K., \& Kitayama, T.\ 2005, \pasj, 57, 611 
\end{thebibliography}

\clearpage

\end{document}